\pdfoutput=1
\documentclass[twocolumn,english,aps,superscriptaddress,prb,longbibliobraphy]{revtex4-1}

\usepackage{babel}
\usepackage{amsmath}
\usepackage{amssymb}
\usepackage{graphicx}

\begin{document}

\title{Probing universal critical scaling with scan-DMRG}

\author{Natalia Chepiga}
\affiliation{Kavli Institute of Nanoscience, Delft University of Technology, Lorentzweg 1, 2628 CJ Delft, The Netherlands}

\date{\today}
\begin{abstract} 
We explore the universal signatures of quantum phase transitions that can be extracted with the density matrix renormalization group (DMRG) algorithm applied to quantum chains with a gradient. We present high-quality data collapses for the order parameter and for the entanglement entropy for three minimal models: transverse-field Ising, three-state Potts and Ashkin-Teller. 
Furthermore, we show that scan-DMRG successfully captures the universal critical scaling when applied across the magnetic Wess-Zumino-Witten and non-magnetic Ising transitions in the frustrated Haldane chain.  In addition,  we report a universal scaling of the lowest excitation energy as a function of a gradient rate.
 Finally, we argue that the scan-DMRG approach has significantly lower computational cost compare to the conventional DMRG protocols to study quantum phase transitions. 
\end{abstract}
\pacs{
75.10.Jm,75.10.Pq,75.40.Mg
}

\maketitle


\section{ Introduction}

Understanding the nature of quantum phase transitions in low-dimensional systems is a central topic of condensed matter physics\cite{giamarchi,tsvelik}. To a large extent modern theory of phase transitions relies on the concept of universality of the critical scaling insensitive to microscopic details of a particular system. This allows to address quantum phase transitions in real materials with simplified lattice models. Despite their simplicity these models very rarely can be solved exactly and otherwise require an advanced numerical techniques. Density matrix renormalization group (DMRG) algorithm\cite{dmrg1,dmrg2,dmrg3,dmrg4} is one of the most accurate and widely used numerical tool for quantum chains. The algorithm relies on the area law stating that entanglement of a low-energy state of local Hamiltonians is capped in one dimension (1D). However, at the critical point the area law is violated and the entanglement entropy along with the computational complexity grow with the system size\cite{CalabreseCardy}. This makes the usage of either finite- or infinite-size DMRG rather challenging in the context of quantum phase transitions\cite{dmrg4,PhysRevLett.132.086503,PhysRevLett.131.226502,3boson,chepiga_dimtrans,PhysRevLett.131.036505}.

 In this paper we present an alternative approach to study phase transitions numerically at much lower computational cost using scan-DMRG.  In this method an external parameter varies along a chain or a cylinder with a fixed gradient.   The scan-DMRG algorithm is traditionally used to probe multiple phases at once - literally, to scan through the phase diagrams \cite{PhysRevB.106.174507,PhysRevLett.120.207203,PhysRevLett.130.116701,jiang2023quantum}. It has also been used to qualitatively distinguish magnetic and non-magnetic domain walls\cite{PhysRevB.100.104426}. In this study, focusing on the situation when two gapped phases are connected through a continuous quantum phase transition, we show how the universal signatures of the phase transition can be extracted with scan-DMRG simulations.
 
Study of quantum chains with linearly varying coupling of field has a long history that roots back to the local density approximation (LDA)\cite{PhysRev.140.A1133} that allows to treat the effect of gradients in the system by assuming that the order parameter is locally uniform. However, LDA cannot be applied near critical points where divergent correlation length is inconsistent with locality constraints. Motivated by the non-homogeneous trapping potentials in cold atoms experiments, the effect of gradients in Luttinger liquids has been intensely studied with the integrable models and by means of conformal field theory \cite{10.21468/SciPostPhys.2.1.002,PhysRevB.96.174301,PhysRevB.83.060414,Bastianello_2020}. The study of gradients that runs through a single transition point between two gapped phases is mainly limited to the transverse-field Ising and XY chains \cite{Zurek_2008,Dziarmaga_2010,2010AdPhy..59.1063D,PhysRevA.82.013630}.   In this paper we will show that the nature of isolated quantum critical point can be captured with gradient chains in a generic case.

Typical numerical analysis of quantum phase transitions relies on the critical scaling of three quantities: the order parameter, the energy gap, and the entanglement entropy. In this paper we will show that in the scan-DMRG all three observables obey the universal scaling. The rest of the paper is organized as follows:
In Section \ref{sec:min} we will present the scan-DMRG results for the three minimal models: Ising, 3-state Potts and Ashkin-Teller describing the transition between the paramagentic disordered phase and phases with spontaneously  broken $\mathbb{Z}_2$, $\mathbb{Z}_3$ and $\mathbb{Z}_4$ symmetries correspondingly. In Section \ref{sec:hald} we will provide computationally more challenging examples of the Ising and Wess-Zumino-Witten (WZW)\cite{difrancesco,AffleckGepner} transitions in a frustrated Haldane chain. In Section \ref{sec:comp} we will discuss the computational gain of the scan-DMRG algorithm. Finally we summarized our results and bring them into a perspective in Section\ref{sec:disc}.

 \section{ Spatial Kibble-Zurek mechanism in minimal models} 
 \label{sec:min}
 \subsection{The idea}
 
 Interpolation from one phase to another along a chain with a fixed gradient can be viewed as a spatial version of the celebrated Kibble-Zurek mechanism\cite{PhysRevB.72.161201,PhysRevLett.95.105701,PhysRevLett.95.245701,kibble_zureck}. In the original formulation of the mechanism the system is dynamically driven through a transition and it turns out that the scaling as a function of sweep rate of certain quantities including the number of topological defects (kinks) and the correlation length are universal for a given type of the transition.
For a constant sweep rate $\tau$ the characteristic time scale is $t\sim \tau ^{\nu/(1+\nu)}$\footnotetext{Here we deal only with conformal transitions and thus set the dynamical critical exponent $z=1$.} and an effective distance to the transition is $\epsilon \sim \frac{t}{\tau} \sim \tau^{\frac{-1}{1+\nu}}$\cite{francuz}. By analogy, we define an effective distance to the transition in the gradient chain:
\begin{equation}
  x \sim (g-g_c)\delta^{\frac{-1}{1+\nu}} \ \ \ \ \ \Rightarrow   \ \ \ \ \ (g-g_c)\sim x\delta^{\frac{1}{1+\nu}}
  \label{eq:eps}
\end{equation}
here $g$ is a parameter varying along the chain (on-site transverse field or coupling constant) and $g_c$ mark the location of the transition in the thermodynamic limit.
The correlation length diverges upon approaching the transition with the critical exponent $\nu$ implying 
\begin{equation}
  \xi\sim (g-g_c)^{-\nu} \sim x^{-\nu}\delta^{\frac{-\nu}{1+\nu}}.
\end{equation}

\subsection{Models}
 
 Ferromagnetic Ising, 3-state Potts and Ashkin-Teller models form an ideal playground to explore quantum phase transitions into $\mathbb{Z}_n$ phases. In the formulation that we use below the location of the transitions is known exactly due to the self-duality of the models. Moreover, the operator content and the set of critical exponents are also well established by conformal field theory (CFT)\cite{difrancesco}.
 
Let us start with  a ferromagnetic transverse field Ising model:
\begin{equation}
  H=-J\sum_iS^x_iS^x_{i+1}-\sum_i h_i S^z_i,
  \label{eq:ising}
\end{equation}
where $S^{x,z}$ are components of the spin-1/2 operator. We set $J=1$ and let the transverse field $h$ to interpolate linearly between $h_\mathrm{st}$ and $h_\mathrm{en}$. In the uniform case the model is critical at $h_c=0.5$ separating ferromagnetic phase with spontaneously broken spin-flip symmetry for $h<h_c$ and a paramagnet for $h>h_c$. The critical point belongs to the 1+1D Ising universality class and is characterized by the central charge $c=1/2$. Critical exponents are $\nu=1$ and $\beta=1/8$. In the gradient model we vary a field term $h_i$. 

Ferromagnetic Ising chain defined in the Eq.\ref{eq:ising} in terms of spin operators $S^{x,z}$ can alternatively be defined in terms of Pauli matrices $\sigma^{x,z}$:
\begin{equation}
  H=-J\sum_i\sigma^x_i\sigma^x_{i+1}-\sum_i h_i \sigma^z_i,
  \label{eq:isingpauli}
\end{equation}
in the present case the critical point is located at $h_c=J$.

Secondly, we formulate the ferromagnetic $q$-state Potts model as a generalization of the Ising model of to the case of local Hilbert space $d=q$ defined by the following microscopic Hamiltonian:
\begin{equation}
  H_\mathrm{Potts}=-J\sum_{i=1}^{N-1}\sum_{\mu=1}^q P_i^\mu P_{i+1}^\mu-\sum_{i=1}^Nh_i P_i,
  \label{eq:potts3}
\end{equation}
where $P_i^\mu=|\mu\rangle_{ii}\langle \mu|-1/q$ tends to project the spin at site $i$ along the $\mu$ direction while $P_i=|\eta_0\rangle_{ii}\langle \eta_0|-1/q$ tends to align spins along the direction $|\eta_0\rangle_i=\sum_\mu |\mu \rangle/\sqrt{q}$. Without loss of generality we set $J=1$ and $h_\mathrm{st}\leq h \leq h_\mathrm{en}$. This model is identical to the transverse-field Ising model of Eq.\ref{eq:isingpauli} when $q=2$.
The 3-state Potts model is realized with $q=3$. In the uniform case its critical point is located at $h_c=1$. The central charge is $c=4/5$ and critical exponents are $\nu=5/6$ and $\beta=1/9$. We introduce gradient through the on-site transverse field term $h_i$.

Further generalization of Ising and 3-state Potts model to the four-dimensional local Hilbert allows an additional freedom. The corresponding minimal model is called the Ashkin-Teller model and is controlled by an asymmetry parameter $\lambda$:
\begin{equation}
  H=-\sum_i J_i (R^1_iR^1_{i+1}+R^2_iR^2_{i+1}+\lambda R^3_iR^3_{i+1})-h\sum_i M,
  \label{eq:AT}
\end{equation}
where transverse field operator is defined $$M(\lambda)=\left( \begin{array}{cccc}
0 & 1 & 1 & \lambda\\
1 & 0 & \lambda & 1\\
1 & \lambda & 0  & 1\\
\lambda & 1 & 1 & 0
\end{array} \right).$$
and $R$-matrices are diagonal with entrees: $R^1=\mathrm{diag}([1,1,-1,-1])$; $R^2=\mathrm{diag}([1,-1,1,-1])$; $R^3=\mathrm{diag}([1,-1,-1,1])$. The parameter $\lambda$ interpolates between two decoupled Ising chains at $\lambda=0$  and the fully-symmetric 4-state Potts point  at $\lambda=1$ that up to a pre-factor resembles Eq.\ref{eq:potts3} with $q=4$.

There is also an alternative definition of the model in terms of Pauli matrices $\sigma^{x,z}$ and $\tau^{x,z}$:
\begin{multline}
H_{AT}=-h\sum_{j=1}^N \left(\sigma_j^x+\tau_j^x+\lambda \sigma_j^x\tau_j^x\right)\\
-\sum_{j=1}^{N-1}J_i \left(\sigma_j^z\sigma_{j+1}^z+\tau_j^z\tau_{j+1}^z+\lambda \sigma_j^z\tau_j^z\sigma_{j+1}^z\tau_{j+1}^z\right),
\end{multline}
Formulated this way, one can easily see that the model is self-dual and in the uniform case its critical point is located at $h=J$ for any value of  $0\leq \lambda\leq 1$.

The Ashkin-Teller universality class is a family of universality classes (also know as a {\it weak} universality class) with some critical exponents continuously varying as a function of the control parameter $\lambda$. For instance, the correlation length critical exponent $\nu$ is given by\cite{kohmoto,obrien}:
\begin{equation}
  \nu=\frac{1}{2-\frac{\pi}{2}\left[\mathrm{acos}(-\lambda)\right]^{-1}},
  \label{eq:nuat}
\end{equation}
while the order parameter critical exponent $\beta=\nu/8$. The central charge at the critical point and the scaling dimension $d=\beta/\nu$ are universal for any $\lambda$ and equal to $c=1$ and $d=1/8$\cite{PhysRevB.91.165129} correspondingly. 
As a main example we will use the Ashkin-Teller model with $\lambda=0.6$. This corresponds to $\nu\approx 0.7748$ and $\beta\approx 0.0969$. This point is away from special symmetry points Ising ($\lambda=0$), 4-state Potts ($\lambda=1$) and parafermions ($\lambda=0.75$) and therefore provides a generic view on the problem.


\subsection{ Order parameter}

In all three models we describe the symmetry broken ferromagnetic phase using a local polarization as an order parameter. \footnote{In order to break the symmetry between the components, we add a boundary field that favors one of the local states at the edge, e.g. $O_i=\langle S^z_i\rangle$ as a local order parameter for the Ising transition with $-h_BS_1^z$ boundary field.} According to the  field theory the order parameter is expected to vanish upon approaching the transition as $O\propto (g-g_c)^\beta$.
In Fig.\ref{fig:magnetization}(a) we show the raw data for the gradient version of the transverse field Ising model where the left and the right edges of the chain are in the ferromagnetic and paramagnetic phases correspondingly, the critical point is at $h_c=0.5$. One can see that the gradient $\delta=(h_\mathrm{en}-h_\mathrm{st})/(N-1)$ blurs the power-law decay of the order parameter. The qualitative effect is quite intuitive - the smaller is the gradient step $\delta$ the closer is the curve to the continuous field theory prediction. Relying on Eq.\ref{eq:eps} we expect 
\begin{equation}
O \sim x^\beta \delta^{\frac{\beta}{1+\nu}}.
\label{eq:op}
\end{equation}
 Similar expression has been obtained for a chain where the gradient runs only through one phase starting with the critical point at one of the edges\cite{2009JSMTE..08..007C}. 
Eq.\ref{eq:op} implies that upon re-scaling the axes to
\begin{equation}
x=(g_i-g_c)\delta^\frac{-1}{1+\nu} \ \ \ \ \  \mathrm{and} \ \ \ \ \  \tilde{O}(x)=O_i\delta^\frac{-\beta}{1+\nu}
\label{eq:axes}
\end{equation}
and $g_i\equiv h_i$, one might expect a perfect data collapse in the critical region. We demonstrate this in Fig.\ref{fig:magnetization}(b) where the data presented in (a) are re-scaled assuming the Ising critical exponents $\beta=1/8$ and $\nu=1$. We use three different windows for the transverse field gradient $|h_i-h_c|\leq 0.01,0.05,0.1$ and various system sizes ranging from $N=200$ to $N=10000$ sites.

\begin{widetext}

\centering

 \begin{figure}[h!]
\centering 
\includegraphics[width=0.85\textwidth]{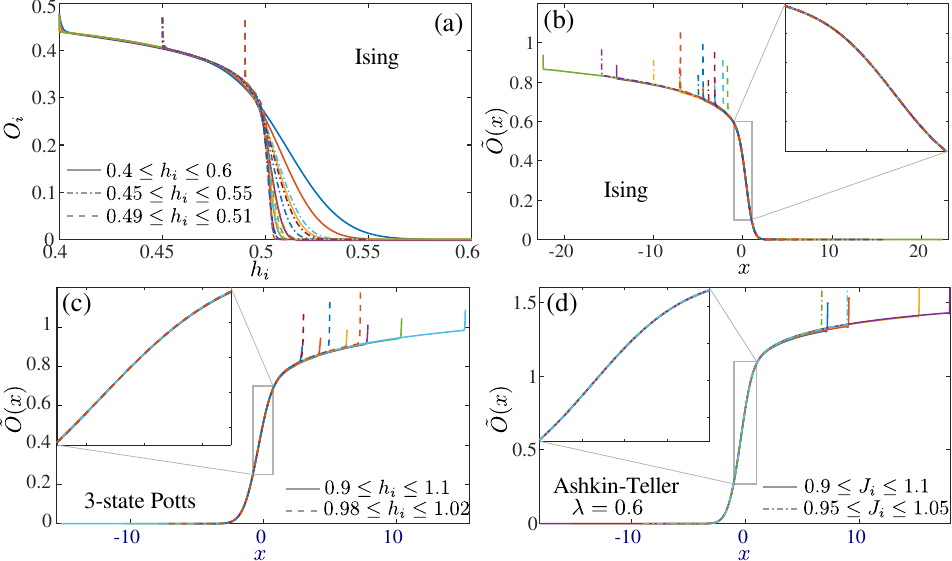}
\caption{Local  order parameter in quantum chains with a fixed gradient. (a) Raw data for local magnetization as a function of on-site transverse field $h_i$ in the Ising model. (b) Same data re-scaled according to Eq.(\ref{eq:axes}): $x=(h_i-h_c)\delta^\frac{-1}{1+\nu}$ and $\tilde{O}(x)=O_i\delta^\frac{-\beta}{1+\nu}$ with Ising critical exponents $\nu=1$ and $\beta=1/8$.  (c)-(d) Local order parameter re-scaled according to Eq.(\ref{eq:axes}) in (c) the gradient 3-state Potts chain with $g_i\equiv h_i$, $\beta=1/9$ and $\nu=5/6$; and (d)  the  Ashkin-Teller gradient model with $\lambda=0.6$, $g_i\equiv J_i$, $\nu\approx0.7748$ and $\beta\approx 0.0969$. The intervals over which the gradients run are indicated in the legends. Insets: zoom over indicated areas.  }
\label{fig:magnetization}
\end{figure}

\end{widetext}

We present similar analysis for the $\mathbb{Z}_3$ transition in the 3-state Potts model defined by Eq.\ref{eq:potts3}. We simulate two windows of the gradient $|h_i-h_c|\leq 0.02,0.1$ using system sizes between  $N=100$ and 2000. 
After re-scaling the axes according to Eq.\ref{eq:axes} with Potts critical exponents  $\beta=1/9$ and $\nu=5/6$ we obtain a spectacular collapse presented in Fig.\ref{fig:magnetization}(c). Magnetization profiles prior to re-scaling are provided in the Appendix \ref{sec:add_AT}.

Finally, in Fig.\ref{fig:magnetization}(d) we show the data collapse for the gradient Ashkin-Teller model defined in Eq.\ref{eq:AT}. Aiming for a generic realization of the Ashkin-Teller transition we focus on $\lambda=0.6$ that corresponds to $\nu\approx0.7748$ and $\beta\approx 0.0969$. In the Appendix \ref{sec:add_AT} we also provide results prior to the re-scaling and  for other values of $\lambda$.  To further stress the universality of our approach in this model we put a gradient in the ferromagnetic interaction ($g_i\equiv J_i$) while keeping the transverse field uniform. As in the case of Ising and 3-state Potts model the collapse presented in Fig.\ref{fig:magnetization}(d) is really spectacular.  

\subsection{ Energy gap } 

Relying on the equivalence of the energy and   length scales at conformal critical points $\Delta\propto \xi^{-1}$ we get:
\begin{equation}
   \Delta\sim \delta^{\frac{\nu}{1+\nu}}
   \label{eq:gap}
\end{equation}
This scaling resembles the form of the energy gap in the Kibble-Zurek mechanism\cite{2010AdPhy..59.1063D}. We compute excitation energies in a gradient chain by targeting multiple states in DMRG\cite{dmrg_chepiga}.

 \begin{figure}[t!]
\centering 
\includegraphics[width=0.5\textwidth]{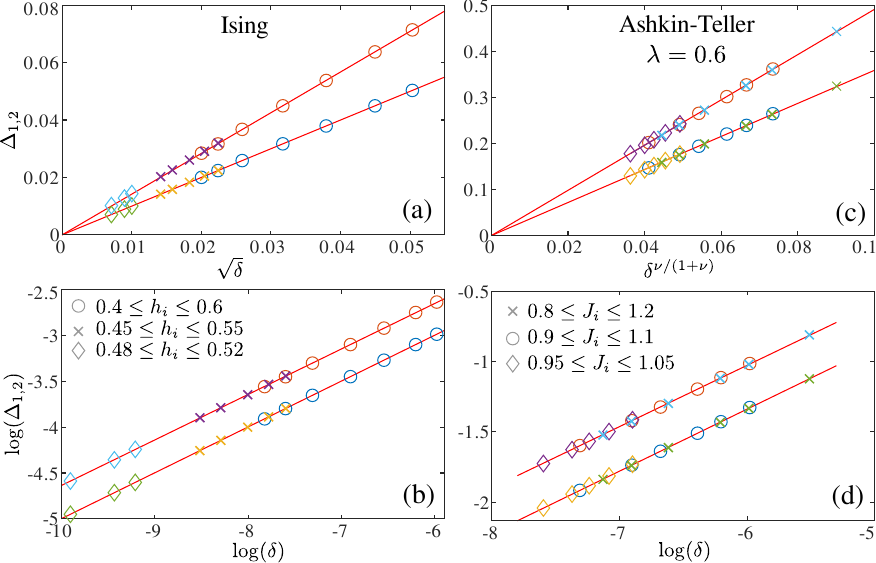}
\caption{Scaling of the excitation energy for the two lowest lying excited states in (a)-(b) Ising  and (c)-(d) Ashkin-Teller models as a function of gradient step $\delta$. Panels (a), (c) present a linear scaling with the inverse of the effective length in agreement with Eq.\ref{eq:gap}. In (b) and (d) we show excitation energy as a function of  a gradient in a log-log scale. Linear fits (red lines) give critical exponents $\nu/(1+\nu)\approx 0.4995$ for the Ising model and $\approx 0.4411$ for the Ashkin-Teller one that are in excellent agreement with theory predictions 0.5 and 0.4366 correspondingly.  For the Ising model we present results for systems ranging between $N=80$ and $N=800$; for the Ashkin-Teller the range is $80\leq N\leq 500$. The windows over which the field/coupling runs are indicated in the legends.  }
\label{fig:gap}
\end{figure}

In Fig.\ref{fig:gap} we present the scaling of the excitation energy for two low-lying excited states: in the panels (a)-(b) we present numerical results for Ising model and  in panels (c),(d) those for the the Ashkin-Teller model with $\lambda=0.6$. The agreement with Eq.\ref{eq:gap} is spectacular: performing a linear fit in the log-log scale gives numerical estimate of the critical exponent $\frac{\nu}{1+\nu}$ agreeing with the theory prediction within $0.1\%$  for the Ising and within $1\%$ for the Ashkin-Teller models.

Interestingly enough, for the Ising model the pre-factor A of the lowest excitation $\Delta_1=A \delta^{\frac{\nu}{1+\nu}}$ is equal to one with a very high precision. This might indicate that {\it for  the lowest excitation} the boundary conditions are approximately equal to polarized at one edge and free at another one, resulting in the conformal tower $\sigma$.

\subsection{ Entanglement entropy }

 Entanglement entropy is extracted numerically using Schmidt values $s_{\alpha}$: $S(i)=-\sum_{\alpha_1}^{D} s_{\alpha}^2\log(s_{\alpha}^2)$, where $i$ is the site index and the size of the subsystem; $D$ is a bond-dimension that controls the accuracy of DMRG simulations. In the thermodynamic limit the entanglement entropy diverges at the critical point. In scan-DMRG the divergence is bounded by both, the finite-size of the (sub)system as well a finite gradient $\delta$. By either increasing the system size or refining the gradient one approaches a continuum limit. This results in a more pronounced peak as shown in Fig.\ref{fig:ent}(a) for the Ising model.

According to the study of non-homogeneous Luttinger liquid\cite{10.21468/SciPostPhys.2.1.002} the gradient $\delta$ produces the $\log$ corrections to the entanglement entropy. We thus define a reduced entanglement entropy as:

\begin{equation}
\tilde{S}(x) \sim S(i) +\frac{c}{12}\log \delta-aC_{i},
\label{eq:ee}
\end{equation}
where $c$ is the central charge; $C_i$ is a nearest-neighbor correlations. The last term is introduced to remove Friedel oscillations (when applicable) from the entanglement entropy profile\cite{PhysRevLett.96.100603,capponi}, this is done by optimizing the  non-universal constant $a$. The re-scaled data as a function of re-scaled distance to the transition $x$ are presented in Fig.\ref{eq:ee}(b) across Ising, (c) across 3-state Potts,  and (d) across Ashkin-Teller transitions characterized by the central charges 1/2, 4/5 and 1 correspondingly. Apart from the boundary effect, the collapse looks perfect in the critical region of all models. We believe that small deviation in the ferromagnetic phase of the Ashkin-Teller model in Fig.\ref{fig:ent}(d) is due to boundary entanglement entropy that comes from the specific fixed  boundary conditions that we use. Further progress of boundary field theory in the context of gradient chains are needed to fully clarify and quantify this effect.

\begin{widetext}
\begin{center}
 \begin{figure}[h!]
\centering 
\includegraphics[width=0.8\textwidth]{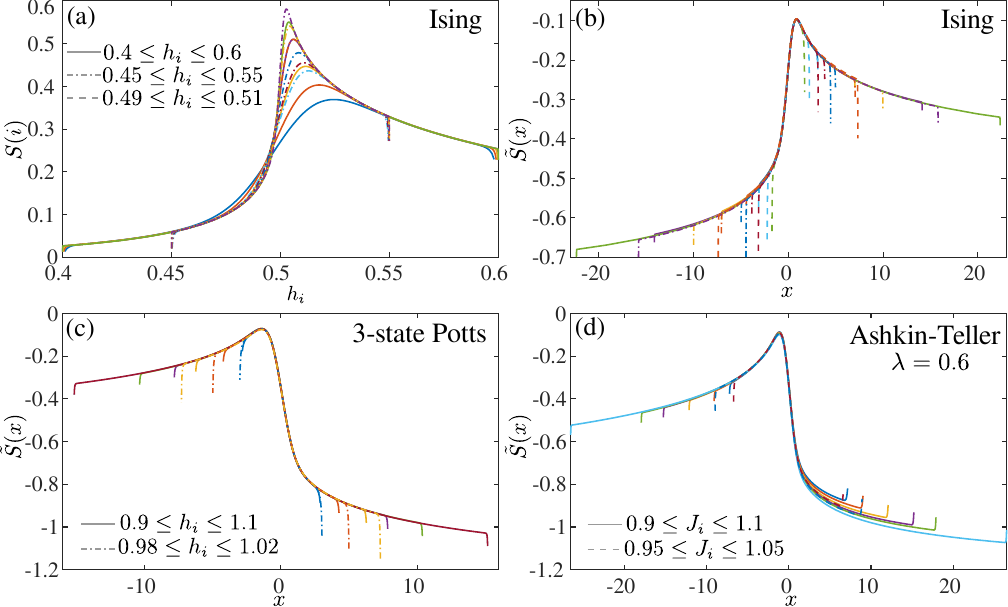}
\caption{Entanglement entropy in a gradient chains. (a) Entanglement entropy of the gradient transverse field Ising model. Peak of entanglement entropy approaches the critical point $h_c=0.5$ upon refining the gradient. (b) Same results re-scaled according to Eq.\ref{eq:ee} and Eq.\ref{eq:eps} with Ising central charge $c=0.5$ leading to a perfect data collapse in the critical region. (c)-(d) Similar data collapce of the entanglement entropy for (c) the 3-state Potts transition with $c=4/5$ and (d) the Ashkin-Teller transition with $c=1$. }
\label{fig:ent}
\end{figure}
\end{center}
\end{widetext}

\pagebreak

\section{Scan-DMRG for the frustrated  Haldane chain}
\label{sec:hald}

In this section we will demonstrate the robustness of the presented scan-DMRG approach beyond the simplest minimal models. As an example we consider dimerization transitions in the frustrated Haldane chain: 1) non-magnetic Ising transition; and 2) magnetic Wess-Zumino-Witten (WZW) SU(2)$_2$ transition. Both have been realized in the spin-1 chain with bilinear-biquadratic interaction\cite{takhtajan,babujian,chepiga_comment} and in the $J_1-J_2-J_3$ model\cite{chepiga_dimtrans} that we will use here and for which the location of the critical points is known with a good accuracy. The $J_1-J_2-J_3$ model is defined by the following microscopic Hamiltonian:

\begin{multline}
  H=J_1\sum_i {\bf S}_i\cdot {\bf S}_{i+1}+J_2\sum_i {\bf S}_i\cdot {\bf S}_{i+2}\\
  +\sum_i J_{3i}\left [( {\bf S}_i\cdot {\bf S}_{i+1})( {\bf S}_{i+1}\cdot {\bf S}_{i+2})+\mathrm{h.c.}\right ],
\end{multline}
where the first two terms describe Heisenberg interactions on nearest- and next-nearest-neighbors; $J_3$ term appears in the next-to-leading order expansion of the two-band Hubbard model and known to produce the dimerized phase with spontaneously broken translation symmetry and a two-fold degenerate ground state\cite{PhysRevLett.108.127202}. Here we set $J_1=1$.

As a local order parameter we use the dimerization $O(i)=|{\bf S}_i\cdot {\bf S}_{i+1}-{\bf S}_{i+1}\cdot {\bf S}_{i+2}|$ that we further re-scale following Eq.\ref{eq:axes}.
For the Ising transition that takes place at $J_2\approx0.7$ and $J_3\approx0.058$\cite{chepiga_dimtrans} we  observe a perfect collapse of the order parameter in Fig.\ref{fig:spin}(a); edge effects are noticeably stronger in the disordered - next-nearest-neighbor Haldane - phase than those that we have observe in the Ising model in Fig.\ref{fig:magnetization}(b).  Collapse of the entanglement entropy presented in Fig.\ref{fig:spin}(c) is very good in the critical region close to $x=0$, but we also see a systematic deviation in the disordered phase. Similar to the results of the Ashkin-Teller model we believe that there might be an additional contribution due to a specific boundary conditions in the next-nearest-neighbor Haldane phase. In the Appendix \ref{sec:s1} we also present these results prior to the re-scaling.

\begin{widetext}
\begin{center}
 \begin{figure}[h!]
\centering 
\includegraphics[width=0.8\textwidth]{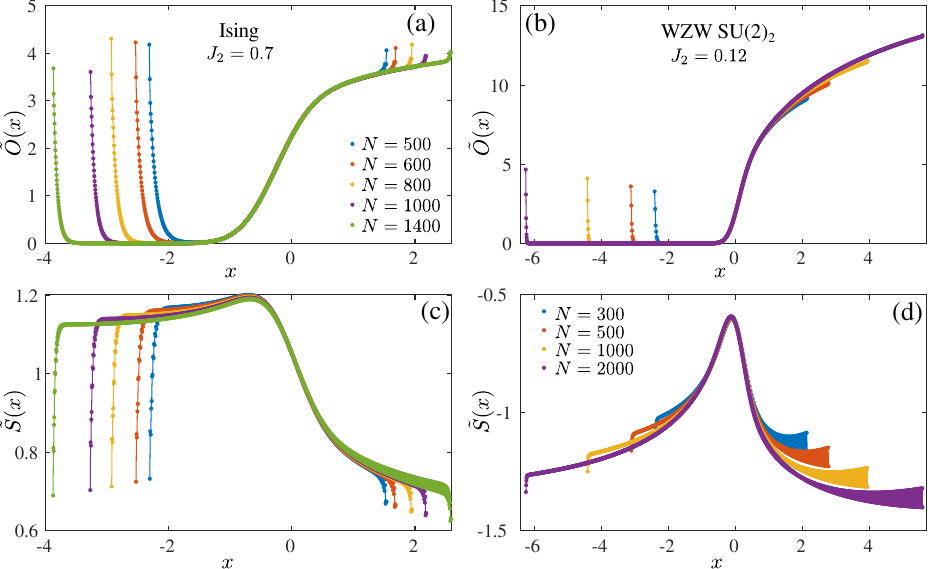}
\caption{Scaling of (a-b) the order parameter and (c-d) the entanglement entropy across (a),(c) non-magnetic Ising transition  and (b),(d) the Wess-Zumino-Witten transition in the frustrated spin-1 chain. For both transition we see an extremely accurate collapse of the order parameter - the dimerization $O(i)=|{\bf S}_i\cdot {\bf S}_{i+1}-{\bf S}_{i+1}\cdot {\bf S}_{i+2}|$}
\label{fig:spin}
\end{figure}
\end{center}
\end{widetext}

To explore the spatial Kibble-Zurek mechanism across the WZW transition we run a gradient through the critical  point $J_2\approx 0.12$, $J_3\approx0.087$ where $\log$-corrections vanish\cite{chepiga_dimtrans}. After the re-scaling the order parameter shows a spectacular collapse with some minor edge effects in the dimerized phase (see Fig.\ref{fig:spin}(b)). Entanglement entropy presented in Fig.\ref{fig:spin}(d) is collapsed in the vicinity of the transitions, though it shows a strong finite-size effects in the domains of the gapped phases far form the critical region. The non-re-scaled data as well as the results for the WZW transition in the presence of log-corrections are reasonably good and are presented in the Appendix \ref{sec:s1}.

\begin{figure}[h!]
\centering 
\includegraphics[width=0.48\textwidth]{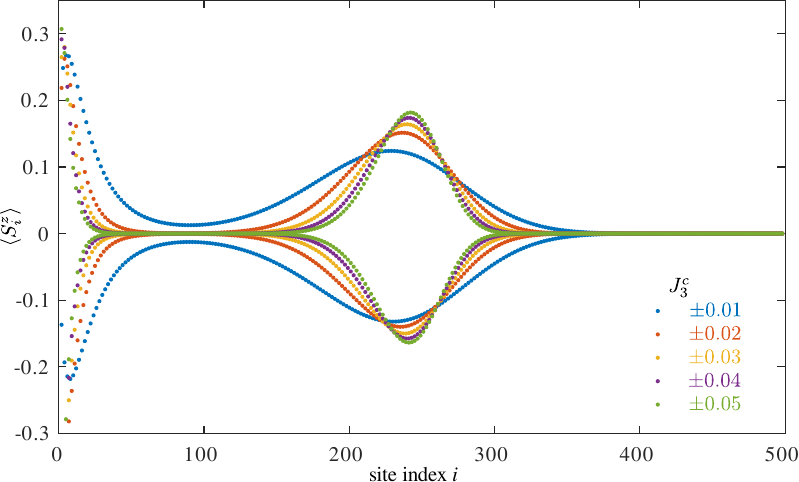}
\caption{Appearance of spinons at the edges of the domain of topologically non-trivial Haldane phase in scan-DMRG that runs through the WZW transition in $J_1-J_2-J_3$ model with $J_2=0.12$. Legend specifies the window of the $J_3$ coupling. These results were obtained with $N=500$. One spin-1/2 is localized at the left edge of the chain. The second spin-1/2 appears as a magnetic domain wall between the Haldane and the dimerized phases. }
\label{fig:spinon}
\end{figure}

With the scan-DMRG one can also very naturally identify the nature of domain walls. For instance, the domain wall between topologically non-trivial Haldane phase and topologically trivial dimerized one hosts a spinon\cite{PhysRevB.94.205112,PhysRevB.101.115138}. This spinon at the boundary of two domains is clearly visible in the local magnetization profile across the WZW transition presented in Fig.\ref{fig:spinon}. By contrast, numerically extracted local magnetization in the gradient spin-1 chain crossing a non-magnetic Ising transition at $J_2=0.7$ never exceeds $10^{-10}$.
 
\section{Computational gain}
\label{sec:comp}
\begin{figure}[h!]
\centering 
\includegraphics[width=0.5\textwidth]{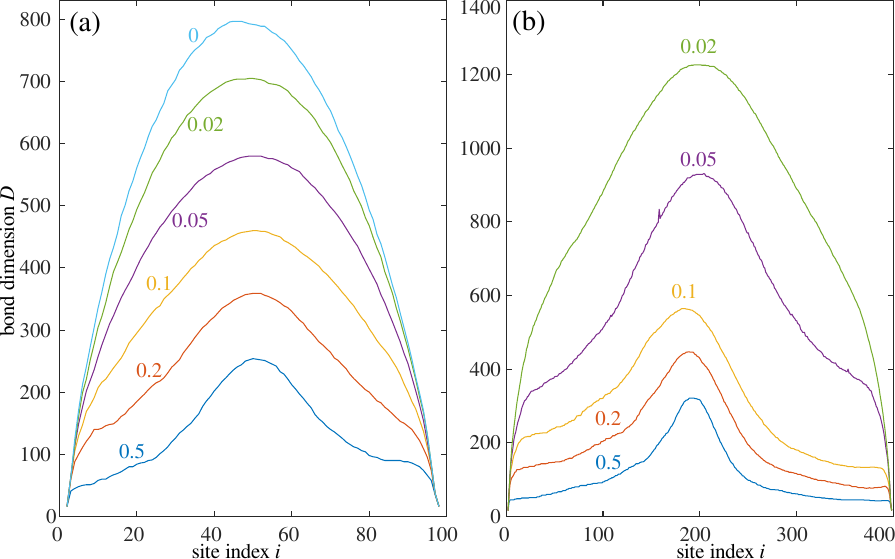}
\caption{Bond dimension along a finite size chain with (a) $N=100$ and (b) $N=400$ sites for the Ashkin-Teller model with $\lambda=0.6$ and with gradient in the coupling around its critical point $J_c=1$. The range of the gradient is indicated for each curve, 0 means uniform case without gradient. We choose asymmetric free-fixed boundary conditions necessary to extract the order parameter with local operators. One can see that the bond dimension and associated complexity of the scan-DMRG that scales as $D^3$ is significantly reduced with the gradient window. Presented bond dimension reflects the number of singular values above $10^{-8}$.  }
\label{fig:bonddim}
\end{figure}

The main advantage of the scan-DMRG technique in the context of quantum criticality is its significantly lower computational cost. Intuitively, this is very natural - the chain almost entirely is inside one of the two gapped phases that obey the area law. Although the area law is violated in the critical region and the entanglement grows it is still capped by the gradient due to spatial Kibble-Zurek mechanism. Let us now quantify possible computational advantage. In the DMRG the entanglement entropy is directly related to the bond dimension $D\propto \exp(S)$, while the leading computational complexity scales as $D^3$. In Fig.\ref{fig:bonddim} we present the bond dimension needed to keep all singular values exceeding $10^{-8}$ for a given system size of (a) $N=100$ and (b) $N=400$ and for a variety coupling ranges around its critical value for the Ashkin-Teller point with $\lambda=0.6$. For a uniform system without a gradient (light blue curve in Fig.\ref{fig:bonddim}(a)) the entanglement along with the bond dimension $D$ are the largest. Slight asymmetry in the dome comes from the asymmetric free-fixed boundary conditions that for consistency we keep the same as for the gradient case. One can see that gradient ranging within $J_i-J_c\in[-0.1,0.1]$ reduces the maximal bond dimension by a factor of $\approx 2$, and thus the complexity is reduced by a factor of 8. Computational gain is even larger for longer chains.

\begin{figure}[h!]
\centering 
\includegraphics[width=0.45\textwidth]{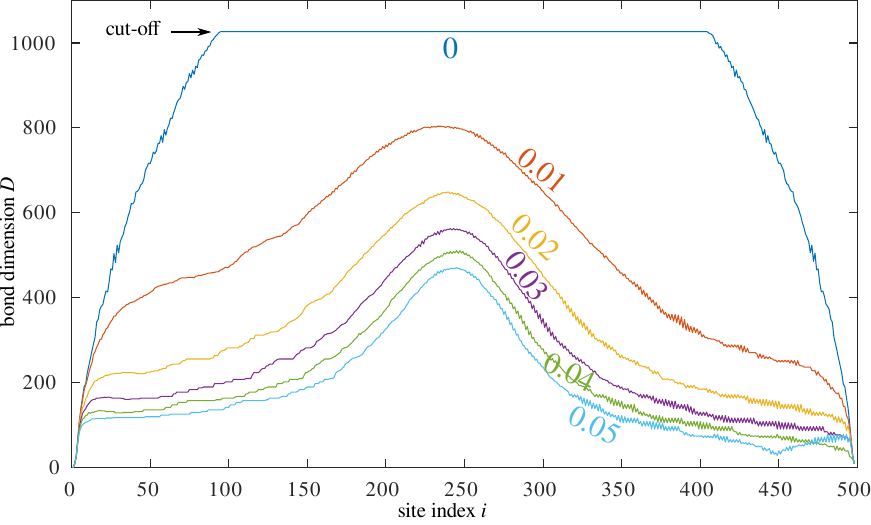}
\caption{Bond dimension along a finite size chain with $N=500$ for the $J_1-J_2-J_3$ model with $J_2=0.12$ and with gradient in three-site coupling $J_3$ around its critical point $J_3^c\approx 0.087$. The range of the gradient is indicated for each curve: 0.05 means the range $0.037\leq J_{3i}\leq 0.127$;  0 means uniform case without a gradient. Presented bond dimension reflects the number of singular values above $10^{-6}$; for the uniform case, we cannot keep sufficient number of states and cut the bond dimension at $D_\mathrm{max}=1027$.  }
\label{fig:bondj1j2j3}
\end{figure}

For the frustrated Haldane chain and WZW transition the computational advantage of scan-DMRG  is even more evident. In Fig.\ref{fig:bondj1j2j3} we present the DMRG bond dimension required to keep all singular values above $10^{-6}$. One can immediately see that in the uniform case the bond dimension grows so fast and actually exceeds our cut-off $D_\mathrm{max}\approx10^3$. By contrast, for the gradient that deviates only by $\mp 0.01$ from the transition the bond dimension hardly reaches $D\approx 800$. In Fig.\ref{fig:spin}(b),(d) we present a scan-DMRG results for the gradient with an approximate range $\pm 0.035$.

\section{Summary and outlook}
\label{sec:disc}

In this paper we have introduced the scan-DMRG algorithm as a numerical tool to investigate universal signatures of quantum phase transitions. We have provided a convincing evidences of the universal scaling of the order parameter, excitation energy, and entanglement entropy across various continuous transitions between gapped phases in quantum chains. We benchmarked our method with three paradigmatic examples of conformal field theory and quantum many-body lattice models - Ising, 3-state Potts and Ashkin-Teller.  Furthermore, we have demonstrated the robustness of the spatial Kibble-Zurek mechanism as a tool to probe quantum phase transitions by applying it to the dimerization transitions in the frustrated Haldane chain. 
We have shown that a convincing data collapse can be produced even when  the location of the critical point is not exact; the interval of the gradient is not centered around the critical point; the translation symmetry broken in one of the phases induces strong Friedel oscillations of the entanglement entropy. 

The main advantage of scan-DMRG as a tool to study quantum phase transitions is the significantly reduced computational cost. Compare to the techniques based on the scaling at the quantum critical points, including, for instance, extraction of the central charge with finite-size entanglement scaling, the scan-DMRG method allow to reduce the computational cost by an order of magnitude. The scan-DMRG also provides a systematic access to the exponents $\beta$ and $\nu$ describing the critical scaling away from the transition. Although with the uniform DMRG algorithm the convergence away from the criticality is typically not an issue, the process of fitting the numerical data in order to extract critical exponents is usually associated with compromises between taking enough points within the critical region (i.e. not too far from the transition) yet to have correlation length and boundary effects sufficiently small compare to the available system sizes (i.e. also not too close to the transition). Typically this requires a dozens of data points on each side of the transition. Scan-DMRG allows an elegant solution to the problem giving a simultaneous access to both sides of the transition, while the quality of the collapse can be assessed even with a very few samples.

Although scan-DMRG appears as a ready-to-use technique for many practical applications, in this work we limit ourselves to the simplest scenario of a direct continuous transition between two gapped phases. This naturally poses the number of relevant questions that are left for future exploration. How and whether  scan-DMRG algorithm can be used for the cases when transition goes through a two-step process with a (possibly narrow) intermediate phase and two distinct transitions? Whether scan-DMRG is suitable to probe continuous transitions where the relevant  order parameter is non-local, including, in particular, the commensurate-incommensurate transitions. Whether scan-DMRG is suitable to distinguish between continuous and first-order transitions\cite{10.21468/SciPostPhys.15.2.061}.
Could scan-DMRG capture the transition when at least one phase is critical, the paradigmatic example could be the Kosterlitz-Thouless transition\cite{Kosterlitz_Thouless}? And if yes, whether there still will be a reasonable advantage in the computational costs? 
We hope that this study will further stimulate analytical and numerical investigation of gradient quantum systems and their applications in the context of cold atom experiments.

 \section{Acknowledgments}
I thank Thierry Giamarchi and Jose Soto Garcia for insightful discussions.
  This research has been supported by Delft Technology Fellowship.  
  Numerical simulations have been performed at the DelftBlue HPC and at the Dutch national e-infrastructure with the support of the SURF Cooperative.
  
\newpage

\begin{appendix}

\section{Additional data for gradient Potts and Ashkin-Teller models}
\label{sec:add_AT}
\begin{figure}[h!]
\centering 
\includegraphics[width=0.4\textwidth]{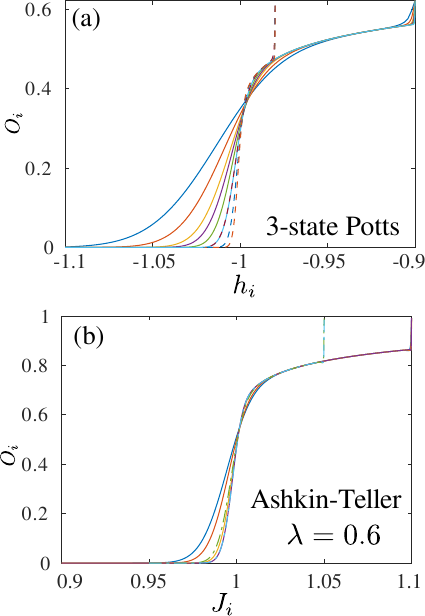}
\caption{ Local magnetization of  (a) the in-homogeneous 3-state Potts with linear gradient in the transverse field and (b) the in-homogeneous Ashkin-Teller model with $\lambda=0.6$ and linear gradient in the couplings constant. As presented in Fig.\ref{fig:magnetization}(c),(d) after re-scaling these sets of  data demonstrate spectacular collapses.}
\label{fig:raw_magn}
\end{figure}

In this section we provide some additional data for the 3-state Potts and Ashkin-Teller models. Firstly,  in Fig.\ref{fig:raw_magn} we present local magnetization profiles without re-scaling as a function of local field $h_i$ (in Fig.\ref{fig:raw_magn}(a)) or a coupling $J_i$ (in Fig.\ref{fig:raw_magn}(b)). The corresponding data collapses are presented in Fig.\ref{fig:magnetization}(c),(d).

\begin{figure}[h!]
\centering 
\includegraphics[width=0.4\textwidth]{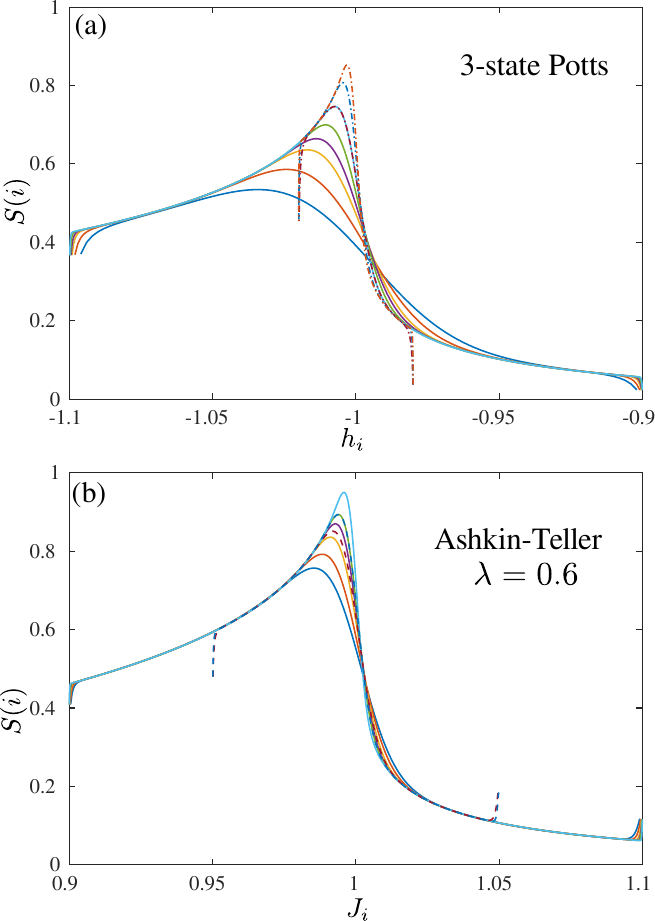}
\caption{ Entanglement entropy of  (a) the in-homogeneous 3-state Potts with linear gradient in the transverse field and (b) the in-homogeneous Ashkin-Teller model with $\lambda=0.6$ and linear gradient in the couplings constant.   Peak of entanglement entropy approaches the critical points upon refining the gradient. The re-scaled data are presented in Fig.\ref{fig:ent}(c),(d).}
\label{fig:raw_ent}
\end{figure}

Secondly, we present the non-rescaled data for the entanglement entropy in the gradient 3-state Potts and Ashkin-Teller models in Fig.\ref{fig:raw_ent}. The re-scaled data are presented in Fig.\ref{fig:ent}(c),(d).
 In both cases the peak of entanglement grows with $1/\delta$. For 3-state Potts model the gradient runs within the window $|h_i-h_c|<0.02$ (dash-dotted line) and $|h_-h_c|<0.1$ (solid lines). We perform simulations on chains with the lengths ranging from 100 and up to 2000 sites. For the Ashkin-Teller model the two parameter windows we use are $|J_i-J_c|<0.05$ and $0.1$ and system sizes ranging between 400 and 2000 sites.

We also present a systematic study of the collapse of the order parameter and the entanglement entropy for various values of $\lambda$ to ensure that our analysis is generic and valid for any critical point of the Ashkin-Teller weak universality class. These results are summarized in Fig.\ref{fig:lambda1}, where in addition to the results for $\lambda=0.6$ presented in the main text we also show the results for $\lambda=0.2$, $0.8$ and $1$. The results for the 4-state Potts point at $\lambda=1$ are of special interest because at this point  the critical Ashkin-Teller theory has log-corrections. From the collapses presented in Fig.\ref{fig:lambda1}(f),(i) we see that for accessible system sizes the effect of these corrections is not visible and the collapses are still very clean.

\begin{widetext}
\begin{center}
 \begin{figure}[h!]
\centering 
\includegraphics[width=0.9\textwidth]{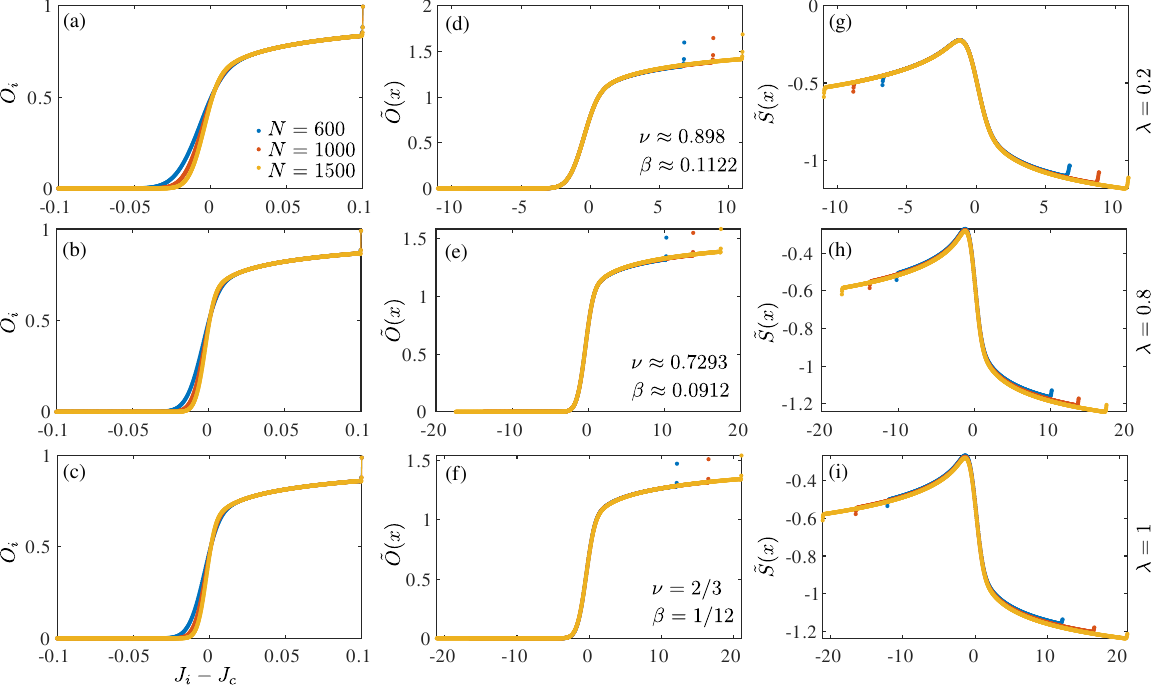}
\caption{Data collapse for various Ashkin-Teller transitions with (top) $\lambda=0.2$, (middle) $\lambda=0.8$ and (bottom) $\lambda=1$ (the 4-state Potts). (a-c) Non-rescaled local order parameter $O_i$ as a function of distance to the critical Ashkin-Teller point $J_c=1$; (d-f) same data re-scaled according to Eq.\ref{eq:axes} with indicated critical exponents. (g-i) Collapse of the re-scaled entanglement entropy.}
\label{fig:lambda1}
\end{figure}
\end{center}
\end{widetext}

\section{Additional data for frustrated spin-1 chain}
\label{sec:s1}

\begin{figure}[h!]
\centering 
\includegraphics[width=0.45\textwidth]{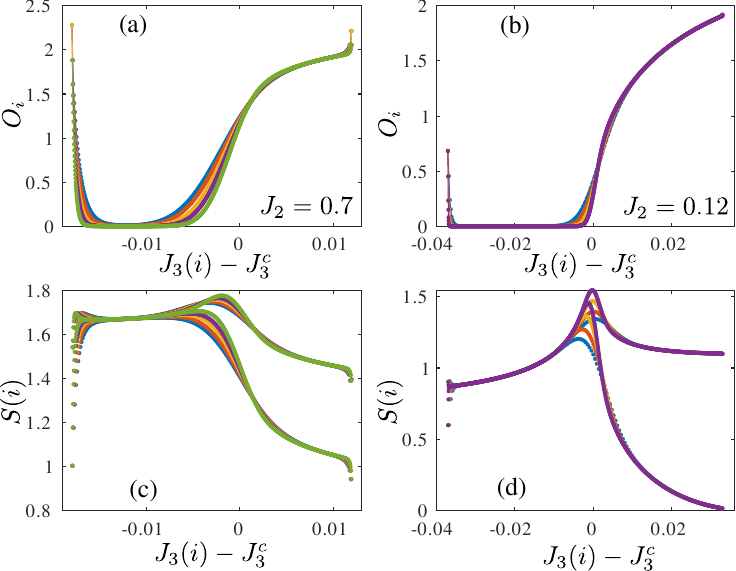}
\caption{ Un-rescaled (a),(b) dimerization and (c),(d) entanglement entropy across (a),(c) Ising  and (b),(d) WZW transitions in spin-1 $J_1-J_2-J_3$ chain\cite{chepiga_dimtrans}. There are strong oscillations in entanglement entropy in the dimerized phase due to broken translation symmetry. (a),(c) For the Ising transition the system sizes range from  500 (blue) to 1400(green). (b),(d) For the WZW transition we show results between $N=300$(blue) and $N=2000$ (purple). The re-scalled data are presented in the Fig.\ref{fig:spin} of the main text.}
\label{fig:raw_spin1}
\end{figure}

In the Fig.\ref{fig:spin} of the main text we presented a data collapse for the Ising and WZW transitions in the $J_1-J_2-J_3$ model. The non-rescaled data are presented in Fig.\ref{fig:raw_spin1}. One can see, in particular, large oscillations of the entanglement entropy in the dimerized phase.

In the main text we probe a special end point of the WZW critical line located at $J_2\approx 0.12$ and $J_3\approx0.087$; at this point log-corrections generically present in the critical theory due to marginal operator vanish.  Here we also present the data collapse for $J_2=0$ where the log-corrections are present. 
We can see that for the entanglement entropy the collapse is still spectacular. For the order parameter we see significant deviations. This is not surprising: the measured effective scaling dimension at $J_2=0$ is $d_\mathrm{eff}\approx0.43$ almost 15\% larger than the field theory prediction $d=3/8$ and the value extracted at the end point\cite{chepiga_dimtrans}. If we allow the critical exponents in the re-scaling functions of Eq.\ref{eq:axes} to deviate within $15\%$ from the CFT values we again recover a perfect collapse. 
Further detailed field-theory analysis is needed to incorporate log-corrections into gradient spin-chains.

 \begin{figure}[h!]
\centering 
\includegraphics[width=0.4\textwidth]{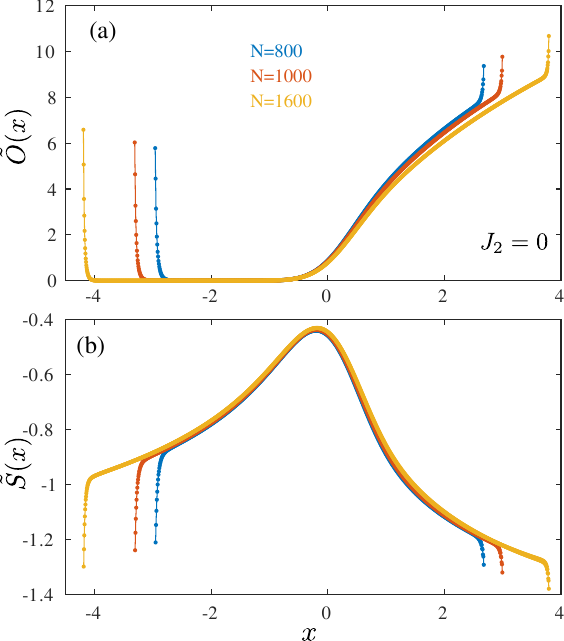}
\caption{ Re-scaled (a) local dimerization and (b) entanglement entropy as a function of the re-scaled three-site coupling constant $J_{3i}$ of $J_1-J_2-J_3$ model with $J_1=1$ and $J_2=0$. The start and end point are fixed to $J_{3,\mathrm{st}}=0.09$ and $J_{3,\mathrm{en}}=0.13$; the critical point is located at $J_3^c\approx0.111$\cite{PhysRevLett.108.127202}. At this point the transition is expected to be in WZW SU(2)$_2$ universality class with some log-corrections.}
\label{fig:wzw0}
\end{figure}

\end{appendix}

\bibliography{bibliography}

\end{document}